\documentclass{llncs}
\usepackage[utf8]{inputenc}
\usepackage{todonotes}
\usepackage{multirow}
\usepackage{booktabs}
\usepackage{tabularx}
\PassOptionsToPackage{hyphens}{url}\usepackage{hyperref}

\usepackage{ amssymb }

\presetkeys{todonotes}{inline}{}

\title{Leveraging Query Resolution and Reading~Comprehension for Conversational~Passage~Retrieval}

\author{Svitlana Vakulenko\inst{1} \and
Nikos Voskarides\inst{1} \and
Zhucheng Tu\inst{2} \and
Shayne Longpre\inst{2}}
\institute{University of Amsterdam  \\ \email{\{s.vakulenko,n.voskarides\}@uva.nl}
\and Apple Inc.}

\begin{document}

\maketitle

\begin{abstract}
    This paper describes the participation of UvA.ILPS group at the TREC CAsT 2020 track. 
    Our passage retrieval pipeline consists of (i) an initial retrieval module that uses BM25, and (ii)  a re-ranking module that combines the score of a BERT ranking model with the score of a machine comprehension model adjusted for passage retrieval.
    An important challenge in conversational passage retrieval is that queries are often under-specified.
    Thus, we perform query resolution, that is, add missing context from the conversation history to the current turn query using QuReTeC, a term classification query resolution model.
    We show that our best automatic and manual runs outperform the corresponding median runs by a large margin.
    
\end{abstract}

\section{Passage Retrieval Pipeline}
\label{sec:pipeline}

\begin{figure}[t]
\includegraphics[width=\textwidth]{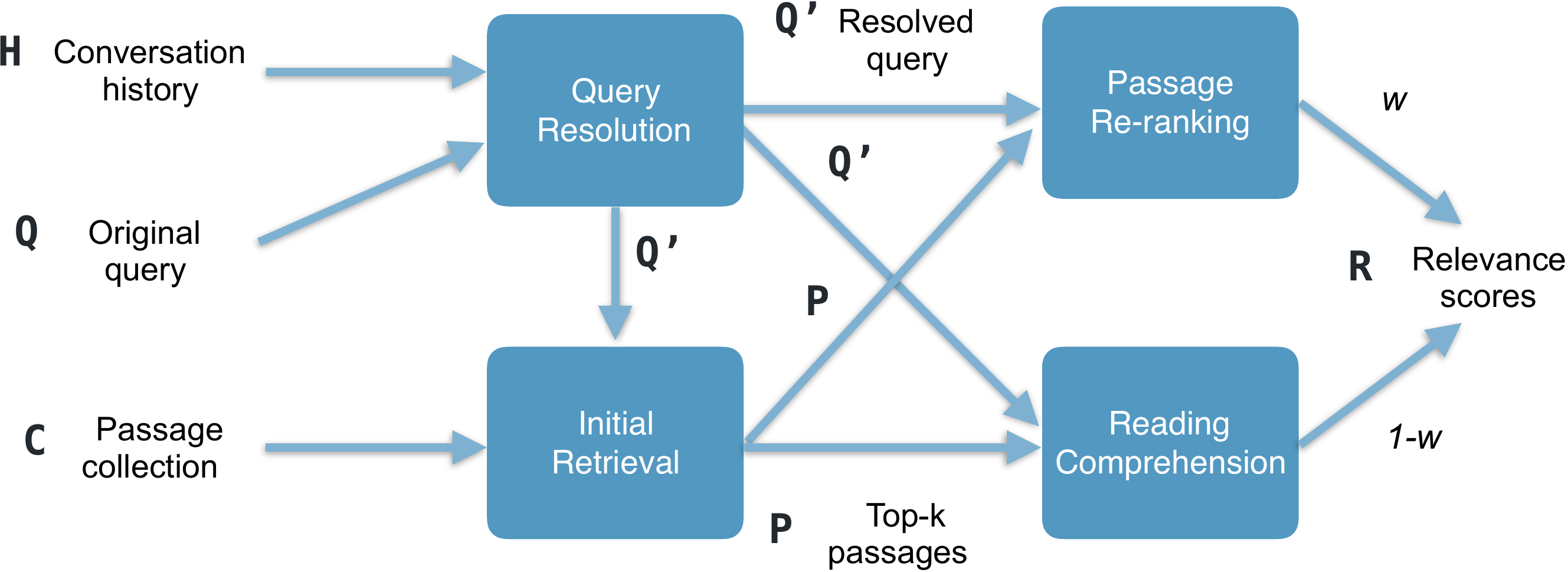}
\caption{Our passage retrieval pipeline.} \label{fig1}
\end{figure}

Our passage retrieval pipeline is shown schematically in Figure~\ref{fig1} and works as follows.
Given the original current turn query $Q$ and the conversation history $H$, we first perform query resolution, that is, add missing context from the $H$ to $Q$ to arrive to the resolved query $Q'$~\cite{DBLP:conf/sigir/VoskaridesLRKR20}.
Next, we perform initial retrieval using $Q'$ to get a list of top-k passages $P$.
Finally, for each passage in $P$, we combine the scores of a reranking module and a reading comprehension module to obtain the final ranked list $R$. %
We describe each module of the pipeline below.

\subsection{Query Resolution}
\label{sec:query-res}

One important challenge in conversational passage retrieval is that the current turn  query is often under-specified.
In order to address this challenge, we perform query resolution, that is, add missing context from the conversation history to the current turn query~\cite{DBLP:conf/sigir/VoskaridesLRKR20}.

We use QuReTeC, a binary term classification query resolution model, which uses BERT to classify each term in the conversation history as relevant or not, and adds the relevant terms to the original current turn query.\footnote{We refer the interested reader to the original paper for more details~\cite{DBLP:conf/sigir/VoskaridesLRKR20}.}
Due to BERT's restrictions on the number of tokens, we cannot include the responses to all the previous turn queries in the conversation history.
Thus, we include (i) all the previous turn queries and (ii) the \emph{automatic} canonical response to the previous turn query only (provided by the track organizers).
We use the QuReTeC model described in~\cite{DBLP:conf/sigir/VoskaridesLRKR20} that was trained on gold standard query resolutions derived from the CANARD dataset~\cite{elgohary2019can}.

\subsection{Initial Retrieval}
\label{sec:initial-retrieval}

We perform initial retrieval using BM25. We tuned the parameters on the MS MARCO passage retrieval dataset ($k_1=0.82, b=0.68$).

\subsection{Re-ranking}
\label{sec:reranking}

Here, we re-rank the original ranking list obtained in the initial retrieval step.
The final ranking score is a weighted average of the Re-ranking (BERT) and Reading Comprehension scores, which we describe below. The  interpolation weight $w$ is optimized on the TREC CAsT 2019 dataset~\cite{DBLP:conf/sigir/0001XKC20}.
\paragraph{Re-ranking (BERT).} We use a BERT model to get a ranking score for each passage as described  in~\cite{nogueira2019passage}. We initialize BERT with \texttt{bert-large} and fine-tuned it on the MS MARCO passage retrieval dataset as described in~\cite{DBLP:journals/corr/abs-2004-14652}.

\paragraph{Reading Comprehension.}
As an additional signal to rank passages we use a reading comprehension model. 
The model is a RoBERTa-Large model trained to predict an answer as a text span in a given passage or ``No Answer'' if the passage does not contain the answer. It is fine-tuned on the MRQA dataset~\cite{fisch2019mrqa}. 
We compute the reading comprehension score as the sum of the predicted start and end span logits: $(l_{start} + l_{end})$.

\section{Runs}
We submitted 3 automatic runs and 1 manual run.
Automatic runs use the raw current turn query, while the manual run uses the manually rewritten current turn query. 
For all runs, we keep the top-\emph{100} ranked passages per query. 

\subsection{Automatic runs}

\begin{itemize}
    \item \texttt{quretecNoRerank}: Uses QuReTeC for query resolution (Section~\ref{sec:query-res}) and the initial retrieval module (Section~\ref{sec:initial-retrieval}), but does not use re-ranking (Section~\ref{sec:reranking}).
    
    \item \texttt{quretecQR}: Uses the whole retrieval pipeline described in Section~\ref{sec:pipeline}.

    \item \texttt{baselineQR}: Uses the whole retrieval pipeline but uses the \emph{automatically} rewritten version of the current turn query provided by the track organizers, instead of QuReTeC. 
    
\end{itemize}

\subsection{Manual run}

\begin{itemize}
    \item \texttt{HumanQR}: Uses the whole retrieval pipeline but uses the \emph{manually} rewritten version of the current turn query provided by the track organizers, instead of QuReTeC.

\end{itemize}

\section{Results}

\begin{table*}[t]
\centering
\caption{Experimental results on the TREC CAsT 2020 dataset. Note that apart from our submitted runs, we also report performance of the Median runs  for reference (\texttt{Median-auto} and \texttt{Median-manual}).}
\label{tab:main}
\begin{tabular}{llrrrrrr}
\toprule
\textbf{Run} & \textbf{Type} & \textbf{NDCG@3} & \textbf{NDCG@5} & \textbf{MAP} & \textbf{MRR} & \textbf{Recall@100}\\
\midrule
\texttt{quretecNoRerank} & Automatic &  0.171 & 0.170 & 0.107 & 0.406 & 0.285  \\
\texttt{Median-auto} & Automatic & 0.225 & 0.220 & 0.145 & - & -  \\
\texttt{baselineQR} & Automatic & 0.319 & 0.302 & 0.158 & 0.556 & 0.266 \\
\texttt{quretecQR} & Automatic &  \textbf{0.340} & \textbf{0.320} & \textbf{0.172} & \textbf{0.589} & \textbf{0.285} \\
\midrule
\texttt{Median-manual} & Manual & 0.317 & 0.303 & 0.201 & - & - \\
\texttt{HumanQR} & Manual & \textbf{0.498} & \textbf{0.472} & \textbf{0.270} & \textbf{0.799} & \textbf{0.408}\\
\bottomrule
\end{tabular}
\end{table*}

Table~\ref{tab:main} shows our experimental results.
First, we observe that  \texttt{quretecNoRerank} underperforms \texttt{Median-auto}, thus highlighting the importance of the re-ranking module.
Also, we observe that \texttt{quretecQR}, the run that uses the whole pipeline, outperforms \texttt{Median-auto} by a large margin and also outperforms \texttt{baselineQR}, on all reported metrics.
This shows the effectiveness of QuReTeC for  query resolution~\cite{DBLP:conf/sigir/VoskaridesLRKR20}.
Moreover, we see that \texttt{quretecQR} is outperformed by  \texttt{humanQR} by a large margin, which highlights the need for future work on the task of query resolution~\cite{vakulenko-2021-comparison}.
Lastly, we observe that our manual run (\texttt{humanQR}) outperforms \texttt{Median-manual}, likely because of better (tuned) retrieval modules.

\section{Analysis}

In this section, we analyze our results using the approach introduced in \cite{DBLP:journals/corr/abs-2010-06835}.

\subsection{Quantitative analysis}

\begin{table}[t]
\centering
\caption{Error analysis when using Original, QuReTeC-resolved or Human queries. 
For a given query group, if NDCG@3\textgreater0 for the query used then we mark it with $\checkmark$, otherwise we mark it with $\times$ (NDCG@3=0). }\label{tab_stats1}
\begin{tabular}{| l | ccc | c | cc |}
\toprule
\multicolumn{1}{|c|}{\multirow{2}{*}{\textbf{Error type}}} & 
\multicolumn{3}{c|}{\textbf{Query}} 
& \multirow{2}{*}{\textbf{\#}} & \multicolumn{2}{c|}{\multirow{2}{*}{\textbf{\%}}}  \\
& \textbf{Original} & \textbf{QuReTeC}-resolved & \textbf{Human} &  &   &  \\
\midrule
\multirow{4}{*}{Ranking error} & $\times$ & $\times$ & $\times$ & 20 & \multicolumn{1}{r|}{9.6} &  \multirow{4}{*}{13.5} \\
& $\checkmark$ & $\times$ & $\times$ & 0 & \multicolumn{1}{r|}{0.0}  &  \\
& $\times$ & $\checkmark$ & $\times$ & 7 & \multicolumn{1}{r|}{3.4}  &  \\
& $\checkmark$ & $\checkmark$ & $\times$ & 1 & \multicolumn{1}{r|}{0.5}  &  \\
\midrule
\multirow{2}{*}{Query resolution error} & $\times$ & $\times$ & $\checkmark$ & 51 & \multicolumn{1}{r|}{24.5} &  \multirow{2}{*}{25.5} \\
& $\checkmark$ & $\times$ & $\checkmark$ & 2 & \multicolumn{1}{r|}{1.0}  &  \\
\midrule
\multirow{2}{*}{No error} & $\times$ & $\checkmark$ & $\checkmark$ & 88 & \multicolumn{1}{r|}{42.2} &
 \multirow{2}{*}{61.0} \\
& $\checkmark$ & $\checkmark$ & $\checkmark$ & 39 & \multicolumn{1}{r|}{18.8}  &  \\
\bottomrule
\end{tabular}
\end{table}

In our pipeline, passage retrieval performance is dependent on the performance of the query resolution module. Thus, we try to estimate the proportion of ranking and query resolution errors separately.
Specifically, we compare passage retrieval performance when using the Original queries, the QuReTeC-resolved queries or Human rewritten queries, and group queries into different types: (i) ranking error, (ii) query resolution error and (iii) no error.
In order to simplify our analysis, we first choose a ranking metric $m$ (e.g., NDCG@3) and a threshold $t$.
We define ranking errors as follows: we assume that Human rewritten queries are always well specified (i.e., they do not need query resolution), and thus poor ranking performance ($m<=t$) when using the Human rewritten queries can be attributed to the ranking modules.
A query resolution error is one for which the Human rewritten query has performance $m>t$, but for which the QuReTeC-resolved query has performance $m<=t$.

Table~\ref{tab_stats1} shows the results of this analysis when using NDCG@3 as the ranking metric $m$ and setting the threshold to $t=0$.
Since we assume that human rewrites are always well specified,  all queries with NDCG@3=0 ($\times$ in column Human) are due to errors in retrieval (13.5\%).
Among the queries for which at least one relevant passage was retrieved in the top-3 (\checkmark in column Human), we see that 61.0\% were correctly resolved by QuReTeC, and 25.5\% were not.
This shows that query resolution for conversational passage retrieval has more room for improvement.
In addition, we observe that $(0+1+2+39)/208\approx$20\% of the queries in the dataset do not need resolution, since when using those we can retrieve at least one relevant passage in the top-3 ($\checkmark$ in column Original).

\begin{table}[t]
\centering
\caption{Error analysis when using Original, QuReTeC-resolved or Human queries. $\checkmark$ indicates that the retrieval performance (NDCG@3 or NDCG@5) reached the threshold indicated in the right columns, and $\times$ indicates that it did not reach the threshold. The numbers correspond to the number of queries in each group.}\label{tab_stats}
\begin{tabular}{| ccc | ccc | ccc |}
\toprule
\multicolumn{3}{|c|}{\textbf{Query}} & \multicolumn{3}{c|}{\textbf{NDCG@3}} & \multicolumn{3}{c|}{\textbf{NDCG@5}} \\
\textbf{Original} & \textbf{QuReTeC}-resolved & \textbf{Human} & \textbf{\textgreater 0} & \textbf{\textgreater{}= 0.5} & \textbf{= 1} & \textbf{\textgreater 0} & \textbf{\textgreater{}= 0.5} & \textbf{= 1} \\
\midrule
$\times$ & $\times$ & $\times$ & 20 & 88 & 185 & 17 & 87 & 196 \\
$\checkmark$ & $\times$ & $\times$ & 0 & 2 & 0 & 0 & 0 & 0 \\
$\times$ & $\checkmark$ & $\times$ & 7 & 3 & 1 & 3 & 3 & 3 \\
$\checkmark$ & $\checkmark$ & $\times$ & 1 & 1 & 0 & 1 & 0 & 0 \\
\midrule
$\times$ & $\times$ & $\checkmark$ & 51 & 42 & 10 & 50 & 48 & 4 \\
$\checkmark$ & $\times$ & $\checkmark$ & 2 & 1 & 0 & 0 & 2 & 0 \\
$\times$ & $\checkmark$ & $\checkmark$ & 88 & 65 & 10 & 87 & 59 & 4 \\
$\checkmark$ & $\checkmark$ & $\checkmark$ & 39 & 6 & 2 & 50 & 9 & 1 \\
\bottomrule
\end{tabular}
\end{table}

\begin{figure}[t]
\centering
\includegraphics[width=\textwidth]{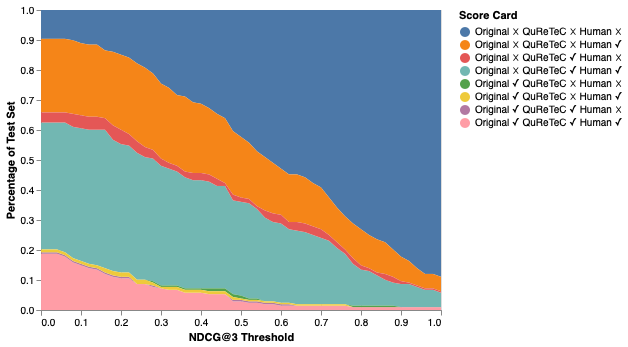}
\caption{Error analysis results using Original, QuReTeC-resolved and Human queries for all thresholds of NDCG@3 for (0; 1] with intervals of 0.02. Passage ranking errors increase as the NDCG threshold increases (blue). The proportion of correct query resolutions (turquoise) is higher than the number of errors produced by QuReTeC (orange). \textit{Best seen in color}.} \label{fig2}
\end{figure}

Table~\ref{tab_stats} shows the same error analysis performed for  different thresholds of NDCG@3 and NDCG@5.
We observe that, as the performance threshold increases, the number of ranking errors increases, which indicates that the passage ranking modules have a lot of room for improvement.
Figure~\ref{fig2} shows the same analysis for NDCG@3, for more thresholds.\footnote{The source code for this analysis that allows to produce the visualisation in Figure~\ref{fig2} from the run files is available at \url{https://github.com/svakulenk0/QRQA}}

\subsection{Qualitative analysis}
\begin{table}[t]
\centering
\caption{Examples where QuReTeC performs worse than Human rewrites.}\label{tab_samples_worse}
\begin{tabularx}{\textwidth}{l@{\hskip .1in}l@{\hskip .1in}Xr}
\toprule
\multicolumn{1}{c}{\textbf{qid}} &  &  & \multicolumn{1}{c}{\textbf{NDCG@3}} \\
\midrule
\multirow{2}{*}{101\_7} & Human & Does the public pay the First Lady of the \textit{United States}?  & 0.864 \\
 & QuReTeC & Do we pay the First Lady? melania trump  & 0 \\ \midrule
\multirow{2}{*}{101\_8} & Human & Does the public pay Ivanka Trump?  & 0.883 \\
 & QuReTeC &What about Ivanka? \textit{melania melanija} trump  & 0 \\ \midrule
\multirow{2}{*}{102\_5} & Human& How much money is owed to social security? &  0.704 \\
 & QuReTeC &How much is owed? \textit{program} social security  & 0 \\ \midrule
\multirow{2}{*}{102\_8} & Human& Can social security be fixed?  & 0.413 \\
 & QuReTeC & Can it be fixed? \textit{checks} social \textit{check} security &  0 \\  \midrule
\multirow{2}{*}{102\_9} & Human & How much of a \textit{tax} increase will keep social security solvent?  & 1.000 \\
 & QuReTeC & How much of an increase? social security &  0 \\
\bottomrule
\end{tabularx}
\end{table}

In order to gain further insights, we sample cases where using the QuReTeC-resolved queries result in a better or worse retrieval performance than when using the Human rewrites.

In Table~\ref{tab_samples_worse} we show examples where QuReTeC performs worse than Human rewrites.
In these cases, QuReTeC either misses relevant tokens or introduces redundant tokens.

\begin{table}[t]
\centering
\caption{Examples where QuReTeC performs better than Human rewrites.}\label{tab_samples_better}
\begin{tabularx}{\textwidth}{l@{\hskip .1in}l@{\hskip .1in}Xr}
\toprule
\multicolumn{1}{c}{\textbf{qid}} &  &  & \multicolumn{1}{c}{\textbf{NDCG@3}} \\
\midrule
\multirow{2}{*}{101\_9} & Human & Does the public pay Jared Kushner?  & 0 \\
 & QuReTeC & And Jared? \textit{ivana donald trump}  & 0.296 \\
 \midrule
\multirow{2}{*}{105\_3} & Human & Why was George Zimmerman acquitted?  & 0 \\
 & QuReTeC & Why was he acquitted? george \textit{trayvon martin} zimmerman  & 0.202 \\ \midrule
\multirow{2}{*}{93\_6} & Human & What support does the franchise provide?  & 0 \\
 & QuReTeC & What support does it provide? \textit{king} franchise  \textit{agreement} \textit{burger}  & 0.521 \\ \midrule
\multirow{2}{*}{98\_7} & Human & Can you show me \textit{vegetarian} recipes with almonds?  & 0 \\
 & QuReTeC & Oh \textit{almonds}? Can you show me recipes with it? \textit{almonds}  & 0.296 \\
\bottomrule
\end{tabularx}
\end{table}
Interestingly, there are also cases in which QuReTeC performs better than Human rewrites (see Table~\ref{tab_samples_better} for examples).
In these examples, QuReTeC introduced tokens from the conversation history that were absent from the  manually rewritten queries but which helped to retrieve relevant passages.

\section{Conclusion}

We presented our participation in the TREC CAsT 2020 track.
We found that our best automatic run that uses QuReTeC for query resolution (\texttt{quretecQR}) outperforms both the automatic median run and the run that uses the rewritten queries provided by the organizers (\texttt{baselineQR}).
In addition, we found that our manual run that uses the human rewrites (\texttt{humanQR}) outperforms our best automatic run (\texttt{quretecQR}), which, alongside with our analysis, highlight the need for further work on the task of query resolution for conversational passage retrieval.

\bibliographystyle{splncs04}
\bibliography{refs}

\end{document}